\documentclass{aa}

\usepackage{graphics}
\usepackage{longtable}
\usepackage{amssymb}

\begin{document}


\title{Analysis of $\alpha$~Centauri AB including seismic constraints}

\author{P.~Eggenberger\inst{1} \and C.~Charbonnel\inst{1,2} \and S.~Talon\inst{3} \and G.~Meynet\inst{1} \and A.~Maeder\inst{1}
        \and F.~Carrier\inst{1} \and G.~Bourban\inst{1}}
   
   \institute{Observatoire de Gen\`eve, CH--1290 Sauverny, Suisse \and
   	      Laboratoire d'Astrophysique de l'OMP, CNRS UMR 5572, 31400 Toulouse, France \and
	      D\'epartement de Physique, Universit\'e de Montr\'eal, Montr\'eal PQ H3C 3J7, Canada}

   \offprints{P. Eggenberger\\
   \email{Patrick.Eggenberger@obs.unige.ch}}

\date{Received / Accepted }
\titlerunning{Analysis of $\alpha$~Centauri AB including seismic constraints}

\abstract{Detailed models of $\alpha$~Cen~A and B based on new seismological data for $\alpha$~Cen~B by Carrier
\& Bourban (\cite{ca03}) have been computed using the Geneva evolution code including atomic diffusion.
Taking into account the numerous observational constraints now available for the $\alpha$~Cen system, we find
a stellar model which is in good agreement with the astrometric, photometric, spectroscopic and asteroseismic data. 
The global parameters of the $\alpha$ Cen system are now firmly constrained to an age of $t=6.52 \pm
0.30$\,Gyr, an initial helium mass fraction $Y_{\mathrm{i}}=0.275 \pm 0.010$ and an initial metallicity
$(Z/X)_{\mathrm{i}}=0.0434 \pm 0.0020$. 
Thanks to these numerous observational constraints, we confirm that the mixing--length
parameter $\alpha$ of the B component is larger than the one of the A component, as already suggested by many authors
(Noels et al. \cite{no91}, Fernandes \& Neuforge \cite{fe95} and Guenther \& Demarque \cite{gu00}):
$\alpha_{\mathrm{B}}$ is about 8\,\% larger than $\alpha_{\mathrm{A}}$ ($\alpha_{\mathrm{A}}=1.83 \pm 0.10$
and $\alpha_{\mathrm{B}}=1.97 \pm 0.10$).
Moreover, we show that asteroseismic measurements enable to determine the radii of both stars with a very high precision 
(errors smaller than 0.3\,\%).
The radii deduced from seismological data are compatible with the new interferometric results of Kervella et al. (\cite{ke03}) even if they
are slightly larger than the interferometric radii (differences smaller than 1\,\%).
}
   
\maketitle

\section{Introduction}
\label{intro}

The $\alpha$~Cen system is the ideal target to test our knowledge of stellar physics in solar--like stars, due to its
proximity which allows precise determination of the fundamental parameters of the two
stars. As a result, numerous theoretical analysis of $\alpha$~Cen~A and B have already been performed.

Flannery \& Ayres (\cite{fl78}) were the first to compute models of the $\alpha$~Cen system. 
They found that stellar models constructed by assuming a solar composition
for $\alpha$~Cen~A and B were not able to reproduce the astrometric and photometric data. They
concluded that the $\alpha$~Cen system is more metal--rich than the Sun ($Z_{\alpha\,\mathrm{Cen}}/Z_{\odot} \sim
2$).

Following the report of a possible detection of p--mode oscillations on $\alpha$~Cen~A (Fossat et al.
\cite{fo84}), Demarque et al. (\cite{de86}) investigated the astrometric and physical properties of the
$\alpha$ Cen system. Firstly, they deduced the masses of the components ($1.09 \pm 0.01$ and 
$0.90 \pm 0.01$\,$M_\odot$ for $\alpha$ Cen A and B respectively) from orbital measurements 
and ground--based parallax. Secondly, they computed stellar models of $\alpha$~Cen~A in order to confront
theoretical predicted p--mode frequencies to observed ones. They concluded that the observed frequencies
were inconsistent with p--mode spectra constructed from standard theoretical models.

Noels et al. (\cite{no91}) introduced a calibration procedure which is based on the simple principle that the
four observables (two luminosities and two effective temperatures) will determine the four unknowns of the
modelisation ($Z$, $Y$, age and mixing--length parameter $\alpha$, which is assumed to be identical 
for both stars). Using the masses deduced by Demarque et al. (\cite{de86}), they found $Z=0.04$,
$Y=0.32$, $\alpha=1.6$ and an age of $5$\,Gyr.  
 
Contrary to Noels et al. (\cite{no91}), Edmonds et al. (\cite{ed92}) relaxed the hypothesis of a unique
mixing--length parameter for both stars by considering the observed metallicity as an additional
constraint.
They were also the first to include the effects of helium diffusion. They found $Y=0.300 \pm 0.005$, an age
of $4.6 \pm 0.4$\,Gyr, and a mixing--length parameter for $\alpha$~Cen~A slightly smaller than for
$\alpha$~Cen~B, but did not include an uncertainty analysis to determine if this difference was significant.  
Edmonds et al. also computed the theoretical p--mode frequencies of $\alpha$~Cen~A and B; they predicted
a mean large spacing (for $\ell=0$) of $108$\,$\mu$Hz and $179$\,$\mu$Hz and a mean small spacing of $6.2$\,$\mu$Hz and $12.6$\,$\mu$Hz for
$\alpha$~Cen~A and B respectively. 

Neuforge (\cite{ne93}) revisited the study by Noels et al. (\cite{no91}) using OPAL opacities. She obtained 
a solution which favours a high value for the metallicity ($Z_{\alpha\,\mathrm{Cen}}/Z_{\odot} \sim
2$) and a unique convection parameter for both stars.

The $\alpha$~Cen system was also studied in order to test the modelisation of convection. Lydon et al.
(\cite{ly93}) constructed a series of models of $\alpha$~Cen~A and B for the purpose of testing the effects of
convection modeling both by means of the mixing--length theory and by means of parametrization of energy fluxes
based upon numerical simulations of turbulent compressible convection. They found that their formulation of
convection produced models with theoretical radii compatible with the observed ones. They were thus able to
correctly modelize a star using a formulation of convection which does not include an adjustable free parameter
to determine the radius.

Fernandes \& Neuforge (\cite{fe95}) also studied the $\alpha$ Cen system to test stellar models based upon the
mixing--length theory of convection and models using the Canuto and Mazzitelli formulation (Canuto \& Mazzitelli
\cite{ca91}, \cite{ca92}). Their calibration suggested that the mixing--length parameter for $\alpha$~Cen~B is
larger that the one for $\alpha$~Cen~A if the mass fraction of heavy elements $Z$ is smaller than 0.038; if
$Z$ is larger than this threshold, the two mixing--length parameters become very similar. However, because of
the uncertainties in the observational constraints, these differences
could not be firmly established.    
 
Kim (\cite{ki99}) calibrated the $\alpha$~Cen system using the observational constraint $[Z/X]$ of
Neuforge--Verheecke \& Magain (\cite{ne97}). His study, which took into account the helium diffusion, gave
an age of about 5.4\,Gyr and showed that both stars may have the same mixing--length ratio ($1.6 \sim 1.7$). Kim
also computed the p--mode frequencies of the $\alpha$~Cen system and predicted a mean large spacing of $104$ and
$171$\,$\mu$Hz for $\alpha$~Cen~A and B respectively.

The different calibrations mentioned previously used the masses derived by Demarque et al. (\cite{de86}): $1.09 \pm 0.01$ and 
$0.90 \pm 0.01$\,$M_\odot$ for $\alpha$ Cen A and B respectively.
Pourbaix et al. (\cite{po99}) performed a simultaneous least--squares adjustment of all
visual and spectroscopic observations of the $\alpha$~Cen system with precise radial velocities measurements.
They derived new consistent values of the orbital parallax, sum of masses, mass ratio and individual masses.
They found masses of $1.16 \pm 0.031$\,$M_\odot$ and $0.97 \pm 0.032$\,$M_\odot$ for $\alpha$~Cen~A and B respectively.
The differences between the masses derived by Pourbaix et al. (\cite{po99}) and the ones obtained by Demarque et al. (\cite{de86})
mainly result from the lower parallax found by Pourbaix et al. ($737.0 \pm 2.6$\,mas
instead of $750.6 \pm 4.6$\,mas for Demarque et al.). In order to investigate the consequences of
these new masses, Pourbaix et al. constructed new models of the $\alpha$~Cen system and found a
significantly smaller age than previous estimates ($2.7$\,Gyr).

Guenther \& Demarque (\cite{gu00}) calculated detailed models of $\alpha$~Cen~A and B based on three
different parallaxes: the parallax from Hipparcos ($742.12 \pm 1.40$\,mas), the parallax of $750.6 \pm 4.6$\,mas obtained by
Demarque et al. (\cite{de86}) and the value of $747.1 \pm 1.2$\,mas determined by S\"oderhjelm (\cite{so99}). Using
the orbital data from Heintz (\cite{he82}) and the mass ratio of Kamper \& Wesselink (\cite{ka78}) they thus
deduced different masses of $\alpha$~Cen~A and B for each parallax: $M_A=1.1238\pm 0.008$\,$M_\odot$ and 
$M_B=0.9344\pm 0.007$\,$M_\odot$ for the parallax from Hipparcos, $M_A=1.0844\pm 0.008$\,$M_\odot$ and 
$M_B=0.9017\pm 0.007$\,$M_\odot$ for the parallax of Demarque et al., and $M_A=1.1015\pm 0.008$\,$M_\odot$ and 
$M_B=0.9159\pm 0.007$\,$M_\odot$ for the parallax determined by S\"oderhjelm. For each pair of masses, Guenther \&
Demarque calculated models of $\alpha$~Cen~A and B including helium and heavy--element diffusion which had to
reproduce the observed effective temperatures,
luminosities (which are slightly different for each pair of masses due to the different parallaxes) and
surface metallicities. They found that self--consistent models of the $\alpha$~Cen system could be produced 
for each pair of masses and thus that the parallax from Demarque et al. as well as the Hipparcos parallax could not
be ruled out, because observational uncertainties in other parameters, such as composition, dominated the uncertainties.
To investigate in details the effect of uncertainties in mass, luminosity, effective temperature, helium aboundance
and metallicity on the models, they chose the models based on the most recent determination of the parallax (the
parallax of S\"oderhjelm). For these models, p--mode frequencies were also computed. 
Guenther \& Demarque found an initial helium mass fraction $Y_{ZAMS}\cong 0.28$ and an age
of the system which depends critically on whether or not $\alpha$~Cen~A has a convective core. If it does, the system has
an age of $7.6 \pm 0.8$\,Gyr. If $\alpha$~Cen~A does not have a convective core (corresponding to $Z_{ZAMS}\lesssim
0.3$), $\alpha$~Cen~A and B are $6.8 \pm 0.8$\,Gyr old. They also found that the mixing--length parameter of
$\alpha$~Cen~A is about 10\,\% smaller that the one of the B component. However, they pointed out that the
effect of composition and surface temperature uncertainties on $\alpha$ is greater than this difference. Concerning
the pulsation analysis of their models, they concluded that the large spacing provides a very precise means of
determining the radius of the star, uncontaminated to any significant degree by uncertainties in the star's
composition. They also pointed out that, because $\alpha$~Cen~A and B have the same age and initial composition,
the small spacing may be useful as an age indicator. They predicted an average large and small spacing of $101\pm
3$\,$\mu$Hz and $4.6\pm 0.4$\,$\mu$Hz for $\alpha$~Cen~A, and $173\pm 6$\,$\mu$Hz 
and $15\pm 1$\,$\mu$Hz for $\alpha$~Cen~B.

Morel et al. (\cite{mo00}) calculated detailed evolutionary models of the $\alpha$~Cen system based on the new masses
of $1.16 \pm 0.031$\,$M_\odot$ and $0.97 \pm 0.032$\,$M_\odot$ for $\alpha$~Cen~A and B respectively (Pourbaix et al. \cite{po99}).
Contrarily to previous works they also chose to use the spectroscopic gravities instead of the luminosity
derived from the photometry, bolometric correction and parallax. The evolutionary code they used included microscopic diffusion. With the mixing--length theory of convection, Morel et
al. found that the $\alpha$~Cen system has an age of $2.71$\,Gyr (in very good agreement with Pourbaix et al.
\cite{po99}), an initial chemical composition $Y_{\mathrm{i}}=0.284$ and
$(Z/X)_{\mathrm{i}}=0.0443$, and
values of the convection parameters which are almost the same for both stars. Including overshooting of
convective cores, they found a larger age of the system ($3.53$\,Gyr), an initial chemical composition
almost the same as without overshooting ($Y_{\mathrm{i}}=0.279$ and
$(Z/X)_{\mathrm{i}}=0.0450$), and
mixing--length parameters almost identical for both components. Morel et al. also performed a calibration of
the $\alpha$~Cen system using the Canuto and Mazzitelli convection theory. They found that this
formulation of the convection changed the results of the
calibration.
Indeed, the age of the $\alpha$~Cen system is larger with the Canuto and Mazzitelli convection theory than with the
mixing--length theory ($4.086$\,Gyr instead of $2.71$\,Gyr). The initial composition derived using the Canuto and
Mazzitelli formulation is $Y_{\mathrm{i}}=0.271$ and $(Z/X)_{\mathrm{i}}=0.0450$. The convection parameters of both
components were also almost identical and close to unity.
Morel et al. also calculated models using the observational constraints adopted by Guenther \& Demarque
(\cite{gu00}). This calibration gave a smaller age ($5.64$\,Gyr) and a larger initial helium mass fraction
($Y_{\mathrm{i}}=0.300$) than obtained by Guenther \& Demarque (\cite{gu00}).
Finally, Morel et al. calculated the p--mode frequencies of their models based on the masses of Pourbaix et al.
(\cite{po99}). They predicted large and small spacings of about $107-108$\,$\mu$Hz and 
$7.5-9.1$\,$\mu$Hz for $\alpha$~Cen~A and $154-157$\,$\mu$Hz and $12$\,$\mu$Hz
for $\alpha$~Cen~B.

The detection and identification of p--modes in $\alpha$~Cen~A by Bouchy \& Carrier (\cite{bo02}) brought strong
additional constraints. These measurements were not fully in agreement with the theoretical calibrations quoted above, which were based 
on non--asteroseismic observables. 
Moreover, Pourbaix et al. (\cite{po02}) studied in detail the masses of the $\alpha$~Cen system. Thanks to the
accurate estimate of the parallax (S\"oderhjelm \cite{so99}) and new radial velocities, they determined very precise
masses of $1.105 \pm 0.0070$\,$M_\odot$ and $0.934 \pm 0.0061$\,$M_\odot$ for $\alpha$~Cen~A and B respectively.

Consequently, new calibrations of the $\alpha$~Cen system were performed by 
Th\'evenin et al. (\cite{th02}) and Thoul et al. (\cite{th03}) in order to find a model reproducing
the asteroseismic observations of the A component as well as the new masses of Pourbaix et al. (\cite{po02}). The analysis of these two theoretical groups led to different 
results. Th\'evenin et al. found a model with an age of $4.85$\,Gyr which is not able to reproduce the astrometric mass of
Pourbaix et al. (\cite{po02}) of $\alpha$~Cen~B, whereas Thoul et al. proposed an older model ($6.41$\,Gyr) which matches
astrometric masses of both stars. The main difference between these two analysis results from the
inclusion of atomic diffusion. Indeed, the evolution code used by Thoul et al. does not include the diffusion
of helium and other heavy elements which cannot be neglected in order to obtain accurate stellar models of
stars in the mass range of $\alpha$~Cen~A and B. On the other hand, Th\'evenin et al. included the atomic diffusion 
in their evolution code and assumed the mixing--length parameter of the two components to be the same. 
Moreover, Th\'evenin et al. used the luminosity derived from the photometry, bolometric correction and parallax
contrary to Thoul et al. who used the spectroscopic surface gravities to constrain the system. 
Both studies
mentioned that, even if the p--mode frequencies of $\alpha$~Cen~A bring strong constraints for the calibration of the $\alpha$~Cen
system, asteroseismic measurements of $\alpha$~Cen~B are needed to unambiguously determine the fundamental parameters of 
$\alpha$~Cen~A and B.

Recently, Carrier \& Bourban (\cite{ca03}) detected solar--like oscillations in $\alpha$~Cen~B with the \textsc{Coralie} 
echelle spectrograph and identified twelve individual frequencies. Thanks to these new asteroseismic measurements, there are
now enough observational constraints to determine an accurate model for the $\alpha$~Cen system. Moreover, 
Kervella et al. (\cite{ke03}) recently measured the angular diameters of $\alpha$~Cen~A and B.
These results enable us to test the consistency between interferometric and asteroseismic observations.  
   
The whole observational constraints available for the $\alpha$~Cen system are presented in Sect.~2.
The physics of the stellar models and the calibration method are described in Sect.~3, while
the results are given in Sect.~4.
Finally Sect.~5 contains the comparison of our results with previous studies and the conclusion is given in Sect.~6.

\section{Observational constraints}
\label{obs}

\subsection{Astrometric data}

The parallax and the orbital elements of the $\alpha$~Cen binary system are difficult to determine accurately due to the high luminosity
and large separation of the stellar components. This results in various values for the parallax published in the literature (see the
introduction and Table~10
of Kervella et al. \cite{ke03}).
In this work we consider the accurate parallax obtained by S\"oderhjelm (\cite{so99}) that combines Hipparcos and ground--based observations, 
$747.1 \pm 1.2$\,mas. 

Pourbaix et al. (\cite{po02}) improved the precision of the orbital
parameters of the $\alpha$~Cen system by adding new radial velocity measurements of $\alpha$~Cen~A and B obtained in the framework of
the Anglo--Australian Planet Search programme as well as in the \textsc{Coralie} programme to those by Endl et al. (\cite{en01}). 
Adopting the parallax of S\"oderhjelm (\cite{so99}), they determined precise masses of $1.105 \pm 0.0070$\,$M_{\odot}$ and $0.934 \pm 0.0061$\,$M_{\odot}$ for $\alpha$~Cen~A and B 
respectively. Note that the mass ratio they determined is compatible with the one obtained by Kamper \& Wesselink
(\cite{ka78}). These new masses constitute the
most recent and accurate values now available for $\alpha$~Cen~A and B. Consequently, we consider them as true
observables which have to be reproduced by a consistent model of the $\alpha$~Cen system.

\subsection{Effective temperatures and chemical composition}
\label{tchim}

Many spectroscopic measurements of both components of the system can be found in the literature. A summary of published values is given
in Table~3 of Morel et al. (\cite{mo00}).   

Concerning the effective temperatures, Th\'evenin et al. (\cite{th02}) and Thoul et al. (\cite{th03}) used the 
same value for $\alpha$~Cen~B. In this work, we considered the same effective temperature of $\alpha$~Cen~B of $5260 \pm 50$\,K. 
For the A component, we note a small discrepancy between the value of Neuforge--Verheecke \& Magain (\cite{ne97})
used by Thoul et al. and the temperature derived by Morel et al. (\cite{mo00}) and used by Th\'evenin et al. We thus adopted an effective 
temperature of $5810 \pm 50$\,K for $\alpha$~Cen~A in order to encompass the intervals of temperatures used by Th\'evenin et al.
and Thoul et al.

For the metallicities, we adopted 
$[\mathrm{Fe/H}]_{\mathrm{A}}=0.22 \pm 0.05$\,dex for $\alpha$~Cen~A and $[\mathrm{Fe/H}]_{\mathrm{B}}=0.24 \pm 0.05$\,dex for 
$\alpha$~Cen~B.
These values lie between the ones used by Th\'evenin et al. and those adopted by Thoul et al, with
larger error boxes which appear to us more realistic in view of the different values found in the literature.

\subsection{Luminosities}
\label{lum}

From 1978 to 1981, both components of the $\alpha$~Cen system have been
measured
in the \textsc{Geneva} photometric system (Golay \cite{golay})
with the photoelectric photometer P7 (Burnet \& Rufener \cite{burnet})
installed on the 40\,cm Swiss telescope in La Silla (ESO, Chile). Six
measurements were
obtained for both stars (see Table~\ref{tab:phot}) using a mask on the
telescope in order 
to reduce the flux of these stars; the photometric reduction procedure is
described 
by Rufener (\cite{rufener1}, \cite{rufener2}).
Combining the mean magnitudes $\langle V_{\mathrm{A}}\rangle = -0.003 \pm 0.006$ and
$\langle V_{\mathrm{B}}\rangle = 1.333 \pm 0.014$\,mag,
the parallax of S\"oderhjelm (\cite{so99}), the solar absolute bolometric magnitude $M_{\mathrm{bol},\,\odot}=4.746$
(Lejeune et al. \cite{le98}) and the
bolometric corrections from Flower's
(\cite{flower}) $BC_{\mathrm{A}} = -0.074 \pm 0.009$ and $BC_{\mathrm{B}} = -0.207 \pm 0.017$\,mag
determined from the effective temperatures, 
luminosities are calculated and have values of $L_{\mathrm{A}} = 1.522 \pm 0.030$\,$L_{\odot}$
and $L_{\mathrm{B}} = 0.503 \pm 0.020$\,$L_{\odot}$.

\begin{table}
\caption{Geneva photometric measurements of $\alpha$~Cen~A and B.
$P$ corresponds to the weight which varies from
0 to 4 according to the quality of the nights (4 for the best nights and 0 for the worst).}
\begin{center}
\begin{tabular}{llll}
\hline
\hline
HJD - 2\,430\,000 & $V_A$ & $V_B$ & $P$ \\ \hline
946.718  & $-$    & 1.296 & 1 \\
946.720  & 0.006  & 1.361 & 1 \\
1056.543 & -0.023 & 1.357 & 1 \\
1063.558 & 0.003  & 1.331 & 1 \\
1070.518 & 0.006  & 1.344 & 1 \\
1324.808 & 0.001  & $-$   & 1 \\
1688.823 & -0.005 & 1.324 & 3 \\ \hline
\end{tabular}
\label{tab:phot}
\end{center}
\end{table}

\subsection{Angular diameters}

Recently Kervella et al. (\cite{ke03}) measured the angular diameters of $\alpha$~Cen~A and B using the VINCI instrument
installed at ESO's VLT Interferometer. They found limb darkened angular diameters $\theta_{A}=8.511 \pm 0.020$\,mas and
$\theta_{B}=6.001 \pm 0.034$\,mas for the A and B component respectively. The radius of each star can then be deduced by using the following 
simple relation between the angular diameter $\theta$ (in mas), the radius of the star $R$ (in $R_{\odot}$) and the parallax $\pi$
(in mas) : $\theta=9.305 \cdot 10^{-3} (2R) \pi$.
By taking into account the error on the parallax of S\"oderhjelm (\cite{so99}), we obtained a radius of $1.224 \pm 0.003$\,$R_{\odot}$ for
$\alpha$~Cen~A and $0.863 \pm 0.005$\,$R_{\odot}$ for $\alpha$~Cen~B.

\begin{table}
\caption{Observational constraints for $\alpha$~Cen~A and B. References: (1) S\"oderhjelm (\cite{so99}), (2) Pourbaix et al. (\cite{po02}),
(3) this paper, (4) derived from the other observational measurements (see text), (5) Kervella et al. (\cite{ke03}),
(6) Bouchy \& Carrier (\cite{bo02}) and (7) Carrier \& Bourban (\cite{ca03}).}
\begin{center}
\begin{tabular}{cccc}
\hline
\hline
 & \multicolumn{1}{c}{$\alpha$~Cen~A} & \multicolumn{1}{c}{$\alpha$~Cen B} & \multicolumn{1}{c}{References} \\ \hline

$\pi$ [mas]& \multicolumn{2}{c}{$747.1 \pm 1.2$} & (1) \\
$M/M_{\odot}$  & $1.105 \pm 0.0070$ & $0.934 \pm 0.0061$ & (2) \\
$V$ [mag] & $-0.003 \pm 0.006$  & $1.333 \pm 0.014$ & (3) \\
$L/L_{\odot}$ & $1.522 \pm 0.030$ & $0.503 \pm 0.020$ & (4) \\
$T_{\mathrm{eff}}$ [K]& $5810 \pm 50$ & $5260 \pm 50$ & (3) \\
$[$Fe/H$]_{\mathrm{s}}$ &  $0.22 \pm 0.05$ & $0.24 \pm 0.05$ & (3) \\
$\theta$ [mas] &  $8.511 \pm 0.020$ & $6.001 \pm 0.034$ & (5)\\
$R/R_{\odot}$ & $1.224 \pm 0.003$ & $0.863 \pm 0.005$ & (5) \\
$\Delta \nu_{0}$ [$\mu$Hz] & $105.5 \pm 0.1$  & $161.1 \pm 0.1$ & (6),(7) \\
$\delta \nu_{02}$ [$\mu$Hz] & $5.6 \pm 0.7$ & $8.7 \pm 0.8$ &  (6),(7) \\
\hline
\label{tab:constraints}
\end{tabular}
\end{center}
\end{table}

\subsection{Asteroseismic constraints}
\label{asc}

Solar--like oscillations in $\alpha$~Cen~A have been detected by Bouchy \& Carrier (\cite{bo02}) with the \textsc{Coralie} echelle spectrograph.
Twenty--eight oscillation frequencies were identified in the power spectrum between 1.8 and 2.9\,mHz with amplitudes in the range 12 to 44\,
cm~s$^{-1}$. The radial orders of these modes with angular degrees $l \leq 2$ lie between 15 and 25. 
According to the asymptotic theory
(Tassoul \cite{ta80}) the oscillation spectrum can be characterized by two frequency separations, the large and the small spacing.
The large frequency spacing $\Delta \nu_{l} \equiv \nu_{n,l}- \nu_{n-1,l}$ corresponds to differences between frequencies of modes with the same angular
degree $l$ and consecutive radial order. For high frequency oscillation modes, this spacing remains approximately constant with a mean value $\Delta \nu_{0}$ 
proportional to the square root of the star's mean density. Thus $\Delta \nu_{0}$ puts strong constraints on the value of the radius of the star and enables
us to test the consistency between interferometric measurements and asteroseismic data.
The small spacing $\delta \nu_{l,l+2} \equiv \nu_{n+1,l}- \nu_{n,l+2}$ is the difference between the frequencies of modes with an angular degree $l$ of
same parity and with consecutive radial order. This small spacing is very sensitive to the structure of the core and mainly to
its hydrogen content, i.e.
to its evolutionary status. Unlike the large spacing, $\delta \nu_{l,l+2}$ can vary sensitively with the frequency. This fact has to be taken into 
account when one compares theoretical value of this mean small spacing with asteroseismic results.  
Bouchy \& Carrier determined the averaged large and small spacing of $\alpha$~Cen~A by a least square fit of the asymptotic relation with their 
twenty--eight oscillations frequencies identified. They found a mean large spacing $\Delta \nu_{0}^{\mathrm{A}}=105.5 \pm 0.1$\,$\mu$Hz and a mean small spacing
between $l=2$ and $l=0$ modes $\delta \nu_{02}^{\mathrm{A}}=5.6 \pm 0.7$\,$\mu$Hz. Note that their fit also gives the value of a third parameter, $\epsilon$,
which is sensitive to the reflexion conditions near the surface of the star. Given that the exact values of the frequencies depend on the details
of the star's atmosphere, where the pulsation is non--adiabatic, we will not use this parameter to constrain our stellar models. Indeed, a linear shift
of a few $\mu$Hz between theoretical and observational frequencies is perfectly acceptable.  

Recently, Carrier \& Bourban (\cite{ca03}) detected solar--like oscillations in $\alpha$~Cen~B with the \textsc{Coralie} echelle spectrograph. They identified
twelve modes between 3 and 4.6\,mHz with amplitudes in the range 8.7 to 13.7\,cm~s$^{-1}$. From these 
twelve frequencies they deduced a large spacing $\Delta \nu_{0}^{\mathrm{B}}=161.1 \pm 0.1$\,$\mu$Hz and a mean small spacing
$\delta \nu_{02}^{\mathrm{B}}=8.7 \pm 0.8$\,$\mu$Hz.

All the observational constraints are listed in Table~\ref{tab:constraints}.

\section{Stellar models}

\subsection{Input Physics}

The stellar evolution code used for these computations is the Geneva code, described several times in the literature 
(see Meynet \& Maeder \cite{mm00} for more details). We used the OPAL opacities, the MHD equation of state (D\"appen et al. 
\cite{da88}; Hummer \& Mihalas \cite{hu88}; Mihalas et al. \cite{mi88}), the NACRE nuclear reaction
rates (Angulo et al. \cite{an99}) and the standard mixing--length formalism for convection. 

Our models have been computed including atomic diffusion on
He, C, N, O, Ne and Mg using the
routines developed for the Geneva--Toulouse version of our code
(see for example Richard et al. \cite{ri96}) recently updated by O. Richard
(private communication).
The diffusion coefficients are computed with the prescription by
Paquette et al. (\cite{pa86}).
We included the diffusion due to the concentration and thermal
gradients, but the radiative acceleration was neglected as it is negligible for the
structure of the low-mass stellar models with extended convective envelopes (Turcotte et al. \cite{tu98}).

\subsection{Calibration method}

Basically, the calibration of a binary system consists in finding the set of stellar modeling parameters which best reproduces all observational 
data available for both stars.
For a given stellar mass the characteristics of a stellar model (luminosity, effective temperature, surface metallicity, frequencies
of oscillation modes, etc.) depend on four modeling parameters: the age of the star (noted $t$ in the following), 
the mixing--length parameter $\alpha \equiv l/H_{\mathrm{p}}$ for convection and two parameters describing the initial chemical composition of the star. For these two parameters,
we chose the initial helium abundance $Y_{\mathrm{i}}$ and the initial ratio between the mass fraction of heavy elements and hydrogen 
$(Z/X)_{\mathrm{i}}$. 
Assuming that this ratio is proportional to the abundance ratio [Fe/H], we can directly relate $(Z/X)$ to [Fe/H]
by using the solar value $(Z/X)_{\odot}=0.0230$ given by Grevesse \& Sauval (\cite{gr98}).
Thus, any characteristic $A$ of a given stellar model has the following formal dependences with respect to modeling parameters :
$A=A(t,\alpha,Y_{\mathrm{i}},(Z/X)_{\mathrm{i}})$.\\
The binary nature of the system provides three constraints: $t_{\mathrm{A}}=t_{\mathrm{B}}$, 
$Y_{\mathrm{i}}^{\mathrm{A}}=Y_{\mathrm{i}}^{\mathrm{B}}$ and $(Z/X)_{\mathrm{i}}^{\mathrm{A}}=(Z/X)_{\mathrm{i}}^{\mathrm{B}}$.
Consequently, we have to determine a set of five modeling parameters ($t$, $\alpha_{\mathrm{A}}$, $\alpha_{\mathrm{B}}$, 
$Y_{\mathrm{i}}$ and $(Z/X)_{\mathrm{i}}$) instead of eight (four for each star). Note that we do not assume that the
mixing--length value is identical
for $\alpha$~Cen~A and B. This assumption was often used in the past 
(Noels et al. \cite{no91})
as an additional constraint needed to close the system, because the only
observed values were the effective temperatures, the masses and the luminosities of both stars.
We no longer need this assumption because we now have enough observational 
data to strongly constrain the models. Moreover, one of the applications of asteroseismology is to determine whether or not
a single mixing--length parameter is applicable to all stars regardless of mass, composition and age.

\subsubsection{Non--asteroseismic constraints}
\label{cal1}

\begin{figure}[htb!]
 \resizebox{\hsize}{!}{\includegraphics{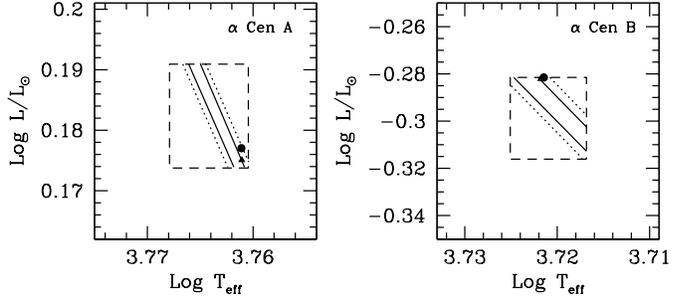}}
  \caption{Observational constraints in the HR diagram for $\alpha$~Cen~A and B. The dashed lines indicate the boxes delimited
  by the observed luminosities and effective temperatures (with their respective 1--sigma errors). The continuous lines
  denote the boxes delimited by the observed radii with their 1--sigma errors, while the dotted line correspond to the same observed
  radii with errors of 2 sigma. The positions in the HR diagram of $\alpha$~Cen~A and B for the M1 model is indicated by the two triangles, while the
  dots correspond to the positions of the M2 model. Note that for $\alpha$~Cen~B the dot and the triangle 
  are partly superimposed, because the M1 model is only slightly hotter and less luminous than the M2
  model.}
  \label{dhr}
\end{figure}

Once assuming that the masses of $\alpha$~Cen~A and B are those determined by Pourbaix et al. (\cite{po02}), the determination of the set of 
modeling parameters ($t$, $\alpha_{\mathrm{A}}$, $\alpha_{\mathrm{B}}$, 
$Y_{\mathrm{i}}$ and $(Z/X)_{\mathrm{i}}$) leading to the best agreement with the observational constraints
is made in two steps. 
First, we only consider non--asteroseismic observations and construct a grid of models with positions of the two components in the HR diagram
in agreement with the observational values of the luminosities, effective temperatures and radii listed in Table~\ref{tab:constraints}.
The boxes in the HR diagram for $\alpha$~Cen~A and B are shown in Fig.~\ref{dhr}. To construct the grid
of models with positions of the two components lying in the observational boxes, we proceed in the following way: for a given chemical composition
(i.e a given set $Y_{\mathrm{i}}$, $(Z/X)_{\mathrm{i}}$) the mixing--length coefficient of each star is adjusted in order to match the observational
position in the HR diagram. 
Note that because of the uncertainties on the effective temperature, different values of the mixing--length parameter
$\alpha$ enable
to match the observational box in the HR diagram. Thus, for each component and for a given initial chemical composition, 
we obtain different models corresponding to different value of $\alpha$ between the smallest value of $\alpha$ needed to match
the observational box (denoted $\alpha_{\mathrm{min}}$) and the largest value ($\alpha_{\mathrm{max}}$). The uncertainty of $50$\,K on
the effective temperature of both stars leads to a typical difference between $\alpha_{\mathrm{min}}$ and $\alpha_{\mathrm{max}}$
of about 0.1.

Once the positions in the HR diagram of the two components agree with the observed values, the surface metallicity of
the two stars is compared to the observed one. If the surface metallicity of one of the component is out of the metallicity intervals listed in 
Table~\ref{tab:constraints}, the models are rejected and the procedure is repeated with another choice of $Y_{\mathrm{i}}$ and $(Z/X)_{\mathrm{i}}$.
Note that the surface metallicities [Fe/H]$_{\mathrm{s}}$ are almost identical for the models 
with the same initial composition and different mixing--length parameters $\alpha$.
Moreover, the [Fe/H]$_{\mathrm{s}}$ of the models are mainly sensitive to $(Z/X)_{\mathrm{i}}$ and less to $Y_{\mathrm{i}}$. 
As a result,
the values of $(Z/X)_{\mathrm{i}}$ are directly constrained by the observed surface metallicities; we found that the models matching the observed
metallicities have $(Z/X)_{\mathrm{i}}$ ranging from about 0.038 to 0.048 (i.e. an initial metallicity [Fe/H]$_{\mathrm{i}}$ between 0.22
and 0.32). 

If the metallicities of both stars agree with the observed values, we then compare their ages. 
All the models of $\alpha$~Cen A with different values of the mixing--length
parameters between
$\alpha_{\mathrm{min}}$ and $\alpha_{\mathrm{max}}$ are considered. For all these models, whose position in the HR diagram as well
as the surface metallicity are in agreement with the observational values, the smallest ($t_{\mathrm{min}}^{\mathrm{A}}$) and the highest ages
($t_{\mathrm{max}}^{\mathrm{A}}$) are determined. Note
that, given the shape of the observational box in the HR diagram (see Fig.~\ref{dhr}), the youngest model has
$\alpha= \alpha_{\mathrm{min}}$ and the oldest has $\alpha= \alpha_{\mathrm{max}}$. 
In the same way, $t_{\mathrm{min}}^{\mathrm{B}}$ and $t_{\mathrm{max}}^{\mathrm{B}}$ are determined for the B component.
If $t_{\mathrm{min}}^{\mathrm{A}} > t_{\mathrm{max}}^{\mathrm{B}}$ or $t_{\mathrm{max}}^{\mathrm{A}} < t_{\mathrm{min}}^{\mathrm{B}}$
the models of $\alpha$~Cen A are not compatible in age with the models of the B component; they are rejected and the procedure is 
repeated with another choice of $Y_{\mathrm{i}}$ and $(Z/X)_{\mathrm{i}}$. 
Otherwise, all the models of $\alpha$~Cen A and B with the same age (with a difference smaller than $0.01$\,Gyr) are considered as models
of the $\alpha$~Cen system which reproduce all the non--asteroseismic constraints. The whole procedure is then 
repeated with a new choice 
of $Y_{\mathrm{i}}$ and $(Z/X)_{\mathrm{i}}$.

In this way, we obtained a grid of models with various sets of modeling parameters ($t$, $\alpha_{\mathrm{A}}$, $\alpha_{\mathrm{B}}$, 
$Y_{\mathrm{i}}$ and $(Z/X)_{\mathrm{i}}$) which satisfied all the non--asteroseismic observational constraints, namely the effective temperatures,
the luminosities, the radii and the surface metallicities. The second step in determining the best model of the binary system $\alpha$~Cen is to consider
the asteroseismic measurements for both stars. 
Note that if no asteroseismic observations were available for the $\alpha$ Cen system,
we would not be able to discriminate between these stellar models. 
Indeed, asteroseismic observations are needed to differentiate models with different internal structures located in the same region of 
the HR diagram, which is absolutely necessary in order to determine the age of the system for instance.

\subsubsection{Pulsation analysis}
\label{cal2}

For each stellar model of the grid constructed as explained above, low--$l$ p--mode frequencies were calculated using the Aarhus adiabatic pulsation
package (Christensen--Dalsgaard \cite{cd97}). Following the observations, frequencies of modes of degree $l \leq 2$ were calculated 
for a radial order $n$ ranging from 15 to 25 for $\alpha$~Cen~A and from 17 to 27 for $\alpha$~Cen B. The determination of the large and
small spacings was carried out exactly as in Bouchy \& Carrier (\cite{bo02}) and Carrier \& Bourban (\cite{ca03}). 
For the models of $\alpha$~Cen~A, we selected the theoretical modes corresponding to the
twenty--eight frequencies identified by Bouchy \& Carrier and fitted the asymptotic relation to obtain the mean values 
$\Delta \nu_0^{\mathrm{A}}$ and $\delta \nu_{02}^{\mathrm{A}}$. In the same way, we selected the theoretical modes of the $\alpha$~Cen B 
models corresponding to the
twelve frequencies identified by Carrier \& Bourban and fitted the asymptotic relation to obtain 
$\Delta \nu_0^{\mathrm{B}}$ and $\delta \nu_{02}^{\mathrm{B}}$. 
 
Once the asteroseismic characteristics of all the relevant models were determined, we performed a $\chi^2$ minimization in order to deduce the 
set of parameters ($t$, $\alpha_{\mathrm{A}}$, $\alpha_{\mathrm{B}}$, $Y_{\mathrm{i}}$, $(Z/X)_{\mathrm{i}}$) leading to the 
best agreement with the observations. Therefore we defined the $\chi^2_{\mathrm{tot}}$ functional
\begin{eqnarray}
\label{eq1}
\chi^2_{\mathrm{tot}} \equiv \sum_{\mathrm{A}, \mathrm{B}} \sum_{i=1}^{6} \left( \frac{C_i^{\mathrm{theo}}-C_i^{\mathrm{obs}}}{\sigma C_i^{\mathrm{obs}}} \right)^2  \; ,
\end{eqnarray}
where the vectors $\mathbf{C}$ contains all the observables for one star:  
\begin{eqnarray}
\nonumber
\mathbf{C} \equiv (L/L_{\odot},T_{\mathrm{eff}},R/R_{\odot},
[\mathrm{Fe/H}]_{\mathrm{s}},\Delta \nu_0,\delta \nu_{02}) \; .   
\end{eqnarray} 
The vector $\mathbf{C}^{\mathrm{theo}}$ contains the theoretical values of these observables for the model to be tested, while 
the values of $\mathbf{C}^{\mathrm{obs}}$ are those
listed in Table~\ref{tab:constraints}. The vector $\mathbf{\sigma C}$ contains the errors on these observations which are also given in
Table~\ref{tab:constraints}.

To better test the agreement between models and asteroseismic observations, we defined a second functional $\chi^2_{\mathrm{astero}}$ which
directly compares individual theoretical frequencies to the observed ones instead of using the mean large and small spacings. As mentioned in 
Sect.~\ref{asc}, a linear shift of a few $\mu$Hz between theoretical and observational frequencies is perfectly acceptable, because the 
exact values of the frequencies depend on the details
of the star's atmosphere, where the pulsation is non--adiabatic. To take this fact into account, we defined the mean value of the differences between
the theoretical and observed frequencies :
\begin{eqnarray}
\nonumber
\langle D_{\nu}\rangle \equiv \frac{1}{N} \sum_{i=1}^N (\nu_i^{\mathrm{theo}}-\nu_i^{\mathrm{obs}}) \; ,
\end{eqnarray} 
where $N$ is the number of observed frequencies ($N_{\mathrm{A}}=28$ for $\alpha$~Cen~A and $N_{\mathrm{B}}=12$ for $\alpha$~Cen B).  
The functional $\chi^2_{\mathrm{astero}}$ can then be defined as

\begin{eqnarray}
\nonumber
\chi^2_{\mathrm{astero}} & \equiv & \frac{1}{N_{\mathrm{A}}} \sum_{i=1}^{N_{\mathrm{A}}} \left(
\frac{\nu_i^{\mathrm{theo}}-\nu_i^{\mathrm{obs}} - \langle D_{\nu}\rangle_{\mathrm{A}}}{\sigma_{\mathrm{A}}} \right)^2 \\
\label{eq2}
& + & \frac{1}{N_{\mathrm{B}}} \sum_{i=1}^{N_{\mathrm{B}}} \left( \frac{\nu_i^{\mathrm{theo}}-\nu_i^{\mathrm{obs}} - \langle
D_{\nu}\rangle_{\mathrm{B}}}{\sigma_{\mathrm{B}}} \right)^2  \; ,
\end{eqnarray} 
where $\sigma_{\mathrm{A}}=0.46$\,$\mu$Hz and $\sigma_{\mathrm{B}}=0.47$\,$\mu$Hz are the errors on the observed frequencies for $\alpha$~Cen~A and B respectively.

The determination of the best set of parameters ($t$, $\alpha_{\mathrm{A}}$, $\alpha_{\mathrm{B}}$, $Y_{\mathrm{i}}$, $(Z/X)_{\mathrm{i}}$) 
was based on the minimization of the functional defined in equation~(\ref{eq1}) which includes four non--asteroseismic
and two asteroseismic observational constraints for each star. Once the model with the smallest $\chi^2_{\mathrm{tot}}$ was determined, we refined the grid in the vicinity of this
preliminary solution in order to find the best solution which minimizes at the same time $\chi^2_{\mathrm{tot}}$ and $\chi^2_{\mathrm{astero}}$.

\section{Results}
\label{res}

\begin{table*}
\caption{Models for $\alpha$~Cen~A and B. The upper part of the table gives the observational constraints used for the
calibration. The middle part of the table presents the modeling parameters with their confidence limits, while the bottom
part presents the global parameters of the two stars. The mean large and small spacings and their respective errors 
are calculated exactly as in Bouchy \& Carrier (\cite{bo02}) and Carrier \& Bourban (\cite{ca03}).}
\begin{center}
\label{tab:res}
\begin{tabular}{c|cc|cc}
\hline
\hline
 & \multicolumn{2}{c}{Model M1} & \multicolumn{2}{|c}{Model M2}  \\
 & \multicolumn{1}{c}{$\alpha$~Cen~A} & \multicolumn{1}{c|}{$\alpha$~Cen B} & \multicolumn{1}{c}{$\alpha$~Cen~A} & \multicolumn{1}{c}{$\alpha$~Cen B} \\ \hline
$M/M_{\odot}$  & $1.105$ & $0.934$ & $1.105$ & $0.934$ \\
$L/L_{\odot}$ & $1.522 \pm 0.030$ & $0.503 \pm 0.020$ & $1.522 \pm 0.030$ & $0.503 \pm 0.020$ \\
$T_{\mathrm{eff}}$ [K]& $5810 \pm 50$ & $5260 \pm 50$ & $5810 \pm 50$ & $5260 \pm 50$ \\
$R/R_{\odot}$ & $1.224 \pm \mathbf{0.003}$ & $0.863 \pm \mathbf{0.005}$ & $1.224 \pm \mathbf{0.006}$ & $0.863 \pm \mathbf{0.010}$ \\
$[$Fe/H$]_{\mathrm{s}}$ &  $0.22 \pm 0.05$ & $0.24 \pm 0.05$ &  $0.22 \pm 0.05$ & $0.24 \pm 0.05$ \\
$\Delta \nu_{0}$ [$\mu$Hz] & $105.5 \pm 0.1$  & $161.1 \pm 0.1$ & $105.5 \pm 0.1$  & $161.1 \pm 0.1$\\
$\delta \nu_{02}$ [$\mu$Hz] & $5.6 \pm 0.7$ & $8.7 \pm 0.8$ & $5.6 \pm 0.7$ & $8.7 \pm 0.8$\\
\hline
$t$ [Gyr] & \multicolumn{2}{c|}{$6.50 \pm 0.20$} & \multicolumn{2}{c}{$6.52 \pm 0.30$} \\
$\alpha$ & $1.83 \pm 0.10$ & $1.99 \pm 0.10$ & $1.83 \pm 0.10$ & $1.97 \pm 0.10$  \\
$Y_{\mathrm{i}}$ & \multicolumn{2}{c|}{$0.275 \pm 0.010$} & \multicolumn{2}{c}{$0.275 \pm 0.010$} \\
$(Z/X)_{\mathrm{i}}$ & \multicolumn{2}{c|}{$0.0435 \pm 0.0020$} & \multicolumn{2}{c}{$0.0434 \pm 0.0020$} \\
\hline
$L/L_{\odot}$ & 1.497 & 0.522 & 1.503 & 0.523 \\
$T_{\mathrm{eff}}$ [K]& $5769$ & $5270$ & $5769$ & $5266$ \\
$R/R_{\odot}$ & $1.227$ & $0.868$ & $1.229$ & $0.870$ \\
$Y_{\mathrm{s}}$ & 0.231 & 0.247 & 0.231 & 0.247 \\
$(Z/X)_{\mathrm{s}}$ & 0.0386 & 0.0402 & 0.0385 & 0.0402 \\
$[$Fe/H$]_{\mathrm{s}}$ &  $0.22$ & $0.24$ &  $0.22$ & $0.24$ \\
$\Delta \nu_{0}$ [$\mu$Hz] & $105.9 \pm 0.1$  & $161.7 \pm 0.1$ & $105.5 \pm 0.1$  & $161.1 \pm 0.1$\\
$\delta \nu_{02}$ [$\mu$Hz] & $4.6 \pm 0.6$ & $10.3 \pm 0.9$ & $4.6 \pm 0.6$ & $10.2 \pm 0.8$\\
\hline
\end{tabular}
\end{center}
\end{table*}

\subsection{Model M1: 1--sigma error on the observed radii}

We first computed a grid of models reproducing all non--asteroseismic constraints within their one--sigma error boxes as described in Sect.~\ref{cal1}.
The positions in the HR diagram for $\alpha$~Cen~A and B are given by their luminosities, effective temperatures and radii 
(see Fig.~\ref{dhr}). Of course, these independent measurements are not fully consistent with each other; for instance, the radius determined by interferometry 
has not exactly the same value as the one deduced from the luminosity and the effective temperature of the star.

\begin{figure}[htb!]
 \resizebox{\hsize}{!}{\includegraphics{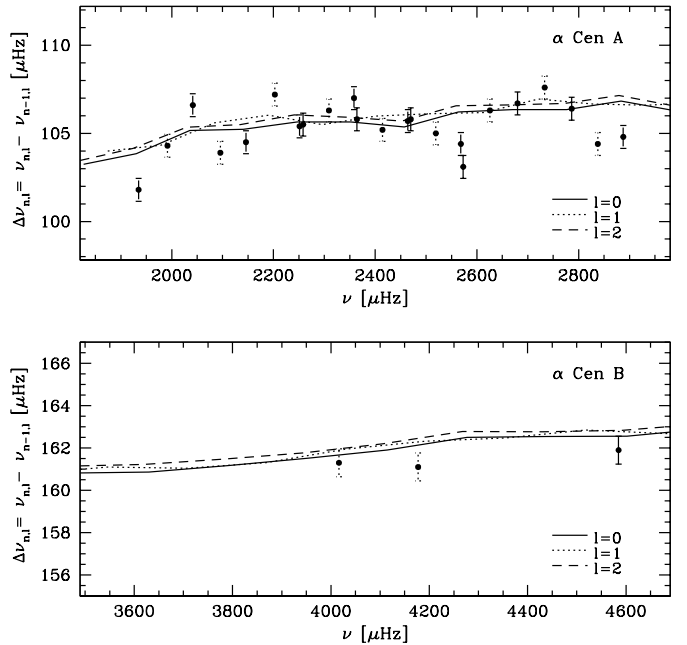}}
  \caption{Large spacing versus frequency for $\alpha$~Cen~A and B for the model with 1--sigma error boxes 
  on the observed radii (model M1). The dots indicate the observed values of the large spacing with an uncertainty on
  individual frequencies estimated to half the time resolution (0.46 and 0.47\,$\mu$Hz for $\alpha$~Cen~A and B respectively).}
  \label{gd_AB_M1}
\end{figure}

\begin{figure}[htb!]
 \resizebox{\hsize}{!}{\includegraphics{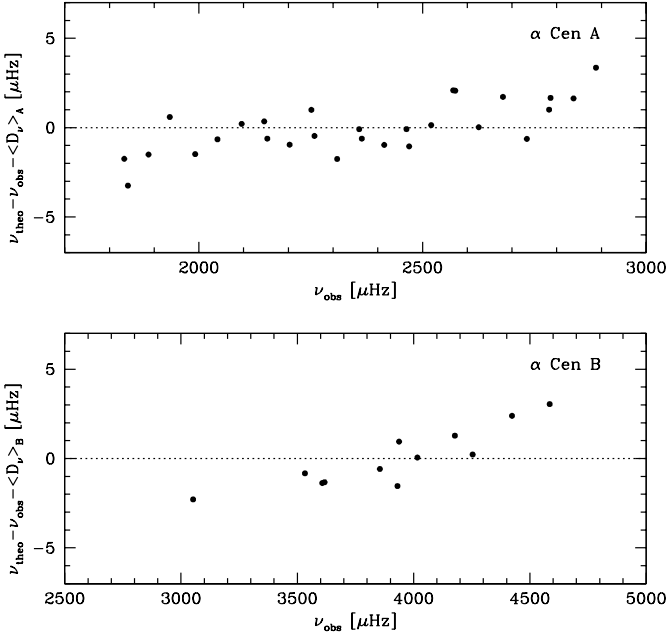}}
  \caption{Differences between calculated and observed frequencies for the model with 1--sigma error boxes on the observed radii (model M1).
  The systematic shifts for $\alpha$~Cen~A and B are $\langle D_{\nu}\rangle_{\mathrm{A}}=-11.5$\,$\mu$Hz and $\langle 
  D_{\nu}\rangle_{\mathrm{B}}=13.7$\,$\mu$Hz 
  (see text for more details).}
  \label{M1theobs}
\end{figure}

Once this grid of models was computed, we performed the $\chi^2$ minimization described above to find the 
set of modeling parameters which best reproduced all observational constraints. In this way, we found the solution
$t=6.50 \pm 0.20$\,Gyr, $\alpha_{\mathrm{A}}=1.83 \pm 0.10$, $\alpha_{\mathrm{B}}=1.99 \pm 0.10$, 
$Y_{\mathrm{i}}=0.275 \pm 0.010$ and $(Z/X)_{\mathrm{i}}=0.0435 \pm 0.0020$. The position in the HR diagram
of this model of $\alpha$~Cen~A and B (denoted model M1 in the following) is given in Fig.~\ref{dhr}. 
The characteristics of this model
are reported in Table~\ref{tab:res}. 
The confidence limits of each modeling parameter given in Table~\ref{tab:res} are estimated as the maximum/minimum values which 
fit the observational constraints when the other calibration parameters are fixed to their medium value.  
The asteroseismic features of this solution are given in Fig.~\ref{gd_AB_M1} and Fig.~\ref{M1theobs}. These two figures
show that this solution is not in complete agreement with the asteroseismic observations. Indeed, the theoretical mean large spacing $\Delta
\nu_0$ of $\alpha$~Cen~A and B are respectively $4 \sigma$ and $6 \sigma$ larger than the observed ones. 
This can be seen in Fig.~\ref{gd_AB_M1} which exhibits the large spacing versus frequency for both stars,
but is more clearly shown in Fig.~\ref{M1theobs} where differences between calculated and observed frequencies are plotted.
In fact, the differences between calculated and observed frequencies increase with frequency. This trend, which is more
pronounced for $\alpha$~Cen B than for $\alpha$~Cen~A, results directly from the too large values of the theoretical 
large spacings. The solution which minimized the $\chi^2$ functional implies that the two stars lie in the upper
border of their respective boxes in radius (see Fig.~\ref{dhr}). Given that the large spacing $\Delta
\nu$ of a star is proportional to the square root of its mean density, we find that the radii observed by interferometry are
slightly smaller that the ones needed to reproduce the asteroseismic observations of $\alpha$~Cen~A and B.

\subsection{Model M2: 2--sigma error on the observed radii}
\label{modM2}

\begin{figure}[htb!]
 \resizebox{\hsize}{!}{\includegraphics{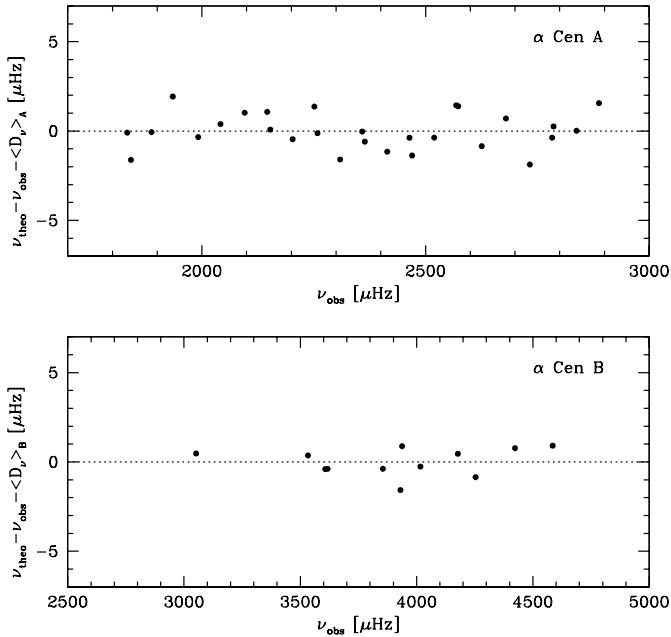}}
  \caption{Differences between calculated and observed frequencies for the model with 2--sigma error boxes on the observed radii (model M2).
  The systematic shifts between theoretical and observed frequencies for $\alpha$~Cen~A and B are $\langle D_{\nu}\rangle_{\mathrm{A}}=-19.0$\,$\mu$Hz
  and $\langle D_{\nu}\rangle_{\mathrm{B}}=0.7$\,$\mu$Hz 
  (see text for more details).}
  \label{M2theobs}
\end{figure}

\begin{figure}[htb!]
 \resizebox{\hsize}{!}{\includegraphics{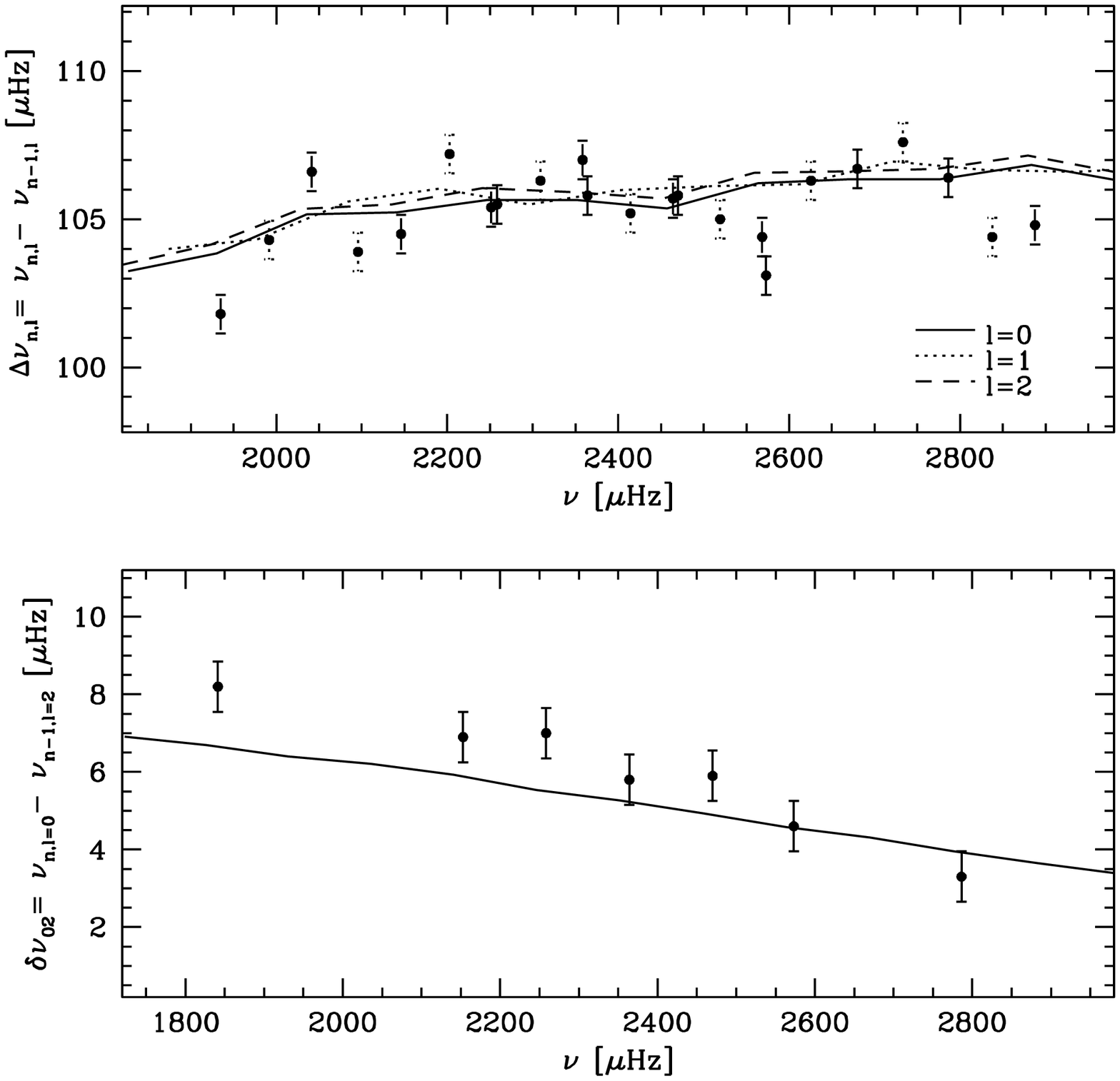}}
  \caption{Large and small spacings versus frequency for $\alpha$~Cen~A for the model with 2--sigma error boxes 
  on the observed radii (model M2). The dots indicate the observed values of the large and small spacings with an uncertainty on
  individual frequencies estimated to half the time resolution (0.46 and 0.47\,$\mu$Hz for $\alpha$~Cen~A and B respectively).}
  \label{gdpt_A_M2}
\end{figure}

\begin{figure}[htb!]
 \resizebox{\hsize}{!}{\includegraphics{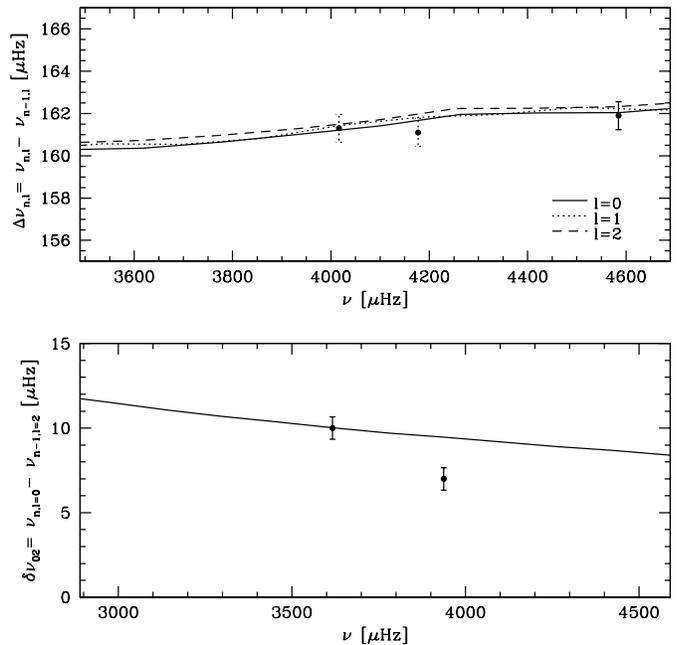}}
  \caption{Large and small spacings versus frequency for $\alpha$~Cen~B for the M2 model.
  The dots indicate the observed values of the large and small spacings with an uncertainty on
  individual frequencies estimated to 0.46 and 0.47\,$\mu$Hz for $\alpha$~Cen~A and B respectively.}
  \label{gdpt_B_M2}
\end{figure}

In order to determine a model of $\alpha$~Cen~A and B which is in better agreement with the asteroseismic measurements, we
decided to repeat the above calibration considering larger values for the errors on the observed radii. We
thus took an error on the observed radii of two sigma instead of the one--sigma error considered previously: $R=1.224 \pm 0.006$\,
$R_{\odot}$ for $\alpha$~Cen~A and $R=0.863 \pm 0.010$\,$R_{\odot}$ for $\alpha$~Cen B.   
For the other observational constraints, we kept the same values as the previous ones. These new
intervals in radius defined a larger region in the HR diagram which is shown in Fig.~\ref{dhr}. With these new constraints on the
radii, we found the solution $t=6.52 \pm 0.30$\,Gyr, $\alpha_{\mathrm{A}}=1.83 \pm 0.10$, $\alpha_{\mathrm{B}}=1.97 \pm 0.10$, 
$Y_{\mathrm{i}}=0.275 \pm 0.010$ and $(Z/X)_{\mathrm{i}}=0.0434 \pm 0.0020$. The position of $\alpha$~Cen~A and B in the HR
diagram for this model (noted model M2 in the following) is denoted by a dot in Fig.~\ref{dhr}. 
The characteristics of this model are given in Table~\ref{tab:res}. Note that the radii of $\alpha$~Cen~A and B for this solution are
only 1.7 and 1.4 sigma larger than the observed values. 
Table~\ref{tab:res} and Fig.~\ref{M2theobs} clearly show that the mean large spacings of the M2
model are in perfect accordance with the observed values. Indeed, the trend visible for the M1 model in Fig.~\ref{M1theobs}
is no longer present in the M2 model. The variations of the large and small spacing with the frequency for $\alpha$~Cen~A and B
are given in Fig.~\ref{gdpt_A_M2} and \ref{gdpt_B_M2}. One can see that the M2 model perfectly reproduces the observed large
spacings. 
It is interesting to notice that asteroseismic observations enable an accurate determination of the radii of both stars.
Indeed, one can see that the M1 model with a radius of $1.227$\,$R_{\odot}$ for $\alpha$ Cen A and 
$0.868$\,$R_{\odot}$ for $\alpha$ Cen B is not compatible with the observed large spacings, while the M2 model with slightly
larger radii ($1.229$\,$R_{\odot}$ for $\alpha$ Cen A and $0.870$\,$R_{\odot}$ for $\alpha$ Cen B) is in perfect agreement 
with the observed large spacings. Consequently, we deduce that the asteroseismic measurements enable to determine the radii of
both stars with a very high precision (errors smaller than $0.3$\,\%).

Concerning the small spacings $\delta \nu_{02}$, we note that the agreement with the observed values is not as
good as for the large spacings. For $\alpha$~Cen~A, the theoretical mean small spacing is compatible with the observational 
measurement but remains slightly smaller than the observed value of 5.6\,$\mu$Hz. For $\alpha$~Cen B, this is exactly the
opposite: the theoretical mean small spacing is higher than the observed one. Since the small spacing $\delta \nu_{02}$ is
very sensitive to the hydrogen content of the core of a main--sequence star, i.e. mainly to its age and its initial chemical
composition, we conclude that the age of about 6.5\,Gyr given by the calibration constitutes the best compromise to reproduce
at the same time the small spacing of $\alpha$~Cen~A and that of $\alpha$~Cen B. Note that the observed mean small
spacing of $\alpha$~Cen B is not strongly constrained since it is deduced from only two points (see Fig.~\ref{gdpt_B_M2}). The
point at low frequency is perfectly reproduced by the model, contrary to the value at higher frequency which is smaller than
predicted by the model. As a result, the theoretical decrease of $\delta \nu_{02}$ with frequency for $\alpha$~Cen B 
is smaller than observed. If this trend for $\alpha$~Cen B can be thought to be artificial given that it is deduced from only two
observational points, it is interesting to mention that this trend is also present for $\alpha$~Cen~A (see Fig.~\ref{gdpt_A_M2}).
Note that for $\alpha$~Cen~A this discrepancy is always present even for stellar models with quite different ages (see Fig.~2 of
Th\'evenin et al. \cite{th02} and Fig.~4 of Thoul et al. \cite{th03}) and remains therefore difficult to explain.

\begin{figure}[htb!]
 \resizebox{\hsize}{!}{\includegraphics{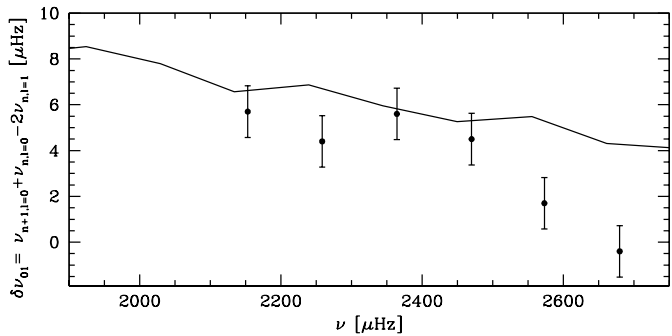}}
  \caption{$\delta \nu_{01}$ spacing versus frequency for $\alpha$~Cen~A for the M2 model. The uncertainty on the observed
  individual frequencies is estimated to 0.46 $\mu$Hz.}
  \label{dnu1_A_M2}
\end{figure}

Previous analysis of the $\alpha$~Cen system made without the asteroseismic constraints on the B component
disagreed on the presence of a convective core in $\alpha$~Cen~A. Th\'evenin et al. (\cite{th02}) found a model with a convective
core, while Thoul et al. (\cite{th03}) proposed a model without a convective core. Our model of $\alpha$~Cen~A does not have a
convective core. However, it is important to note that $\alpha$~Cen~A lies very close to the boundary between models
with and without convective core. Thus small changes in the observational constraints adopted to calibrate the system can lead
to models of $\alpha$~Cen~A with or without a convective core. Moreover, the asteroseismic measurements of $\alpha$~Cen~A are not
precise enough to allow a direct discrimination on this characteristic. As can indeed be seen on
Fig.~\ref{dnu1_A_M2}, the spacing $\delta \nu_{01}=\nu_{n+1,l=0}+ \nu_{n,l=0}- 2\nu_{n,l=1}$
for our best model of $\alpha$~Cen~A is very similar to the spacing $\delta \nu_{01}$ of a model with a convective core (see
Fig.~3 of Th\'evenin et al. \cite{th02}). Therefore it cannot be used to discriminate models with and without a convective
core.   

\begin{figure}[htb!]
 \resizebox{\hsize}{!}{\includegraphics{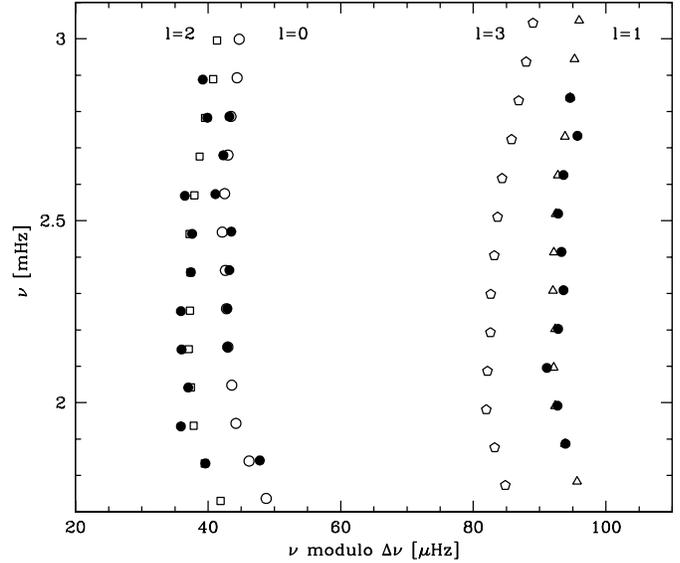}}
  \caption{Echelle diagram of $\alpha$~Cen~A for the best model of the $\alpha$~Cen system (model M2),
with a large spacing $\Delta \nu=105.5$\,$\mu$Hz. Open symbols refer to theoretical
frequencies, while the filled circles correspond to the frequencies observed
by Bouchy \& Carrier (\cite{bo02}). 
Open circles are used for modes with $l=0$, triangles for $l=1$, squares for $l=2$ and pentagons for $l=3$ (see text for more
details).}
  \label{ech_A_M2}
\end{figure}

\begin{figure}[htb!]
 \resizebox{\hsize}{!}{\includegraphics{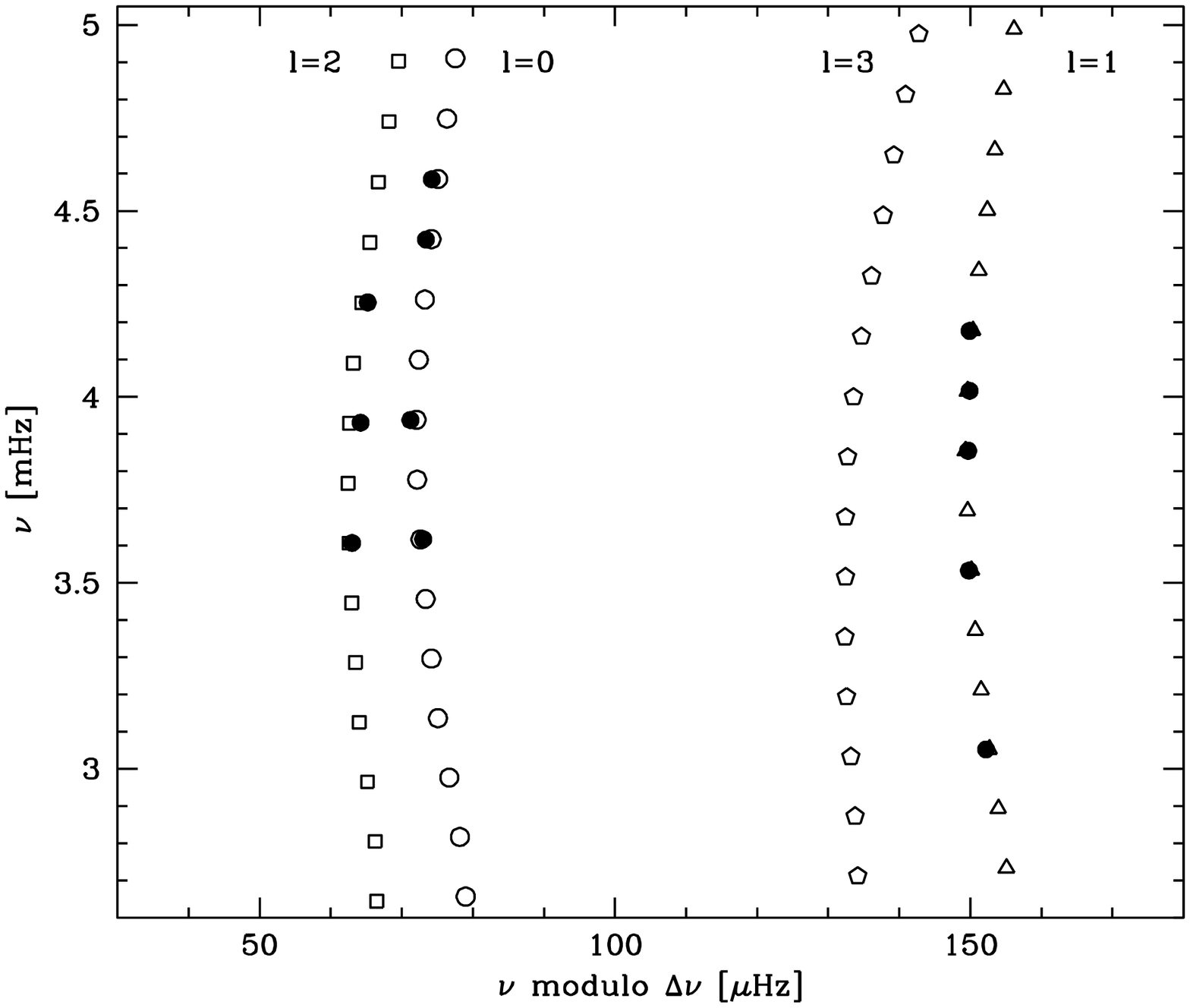}}
  \caption{Echelle diagram of $\alpha$~Cen~B for the best model of the $\alpha$~Cen system (model M2),
with a large spacing $\Delta \nu=161.1$\,$\mu$Hz. Open symbols refer to theoretical
frequencies, while the filled circles correspond to the frequencies observed
by Carrier \& Bourban (\cite{ca03}). 
Open circles are used for modes with $l=0$, triangles for $l=1$, squares for $l=2$ and pentagons for $l=3$ (see text for more
details). }
  \label{ech_B_M2}
\end{figure}

Finally, we compare the theoretical p--mode frequencies of our best model (model M2) to the observed frequencies by
plotting the echelle diagrams of the two stars. Fig.~\ref{ech_A_M2} shows the echelle diagram of $\alpha$~Cen~A, while
that of $\alpha$~Cen B is given in Fig.~\ref{ech_B_M2}. In these two figures, the systematic differences
between theoretical and observed frequencies ($\langle D_{\nu}\rangle_{\mathrm{A}}=-19.0$ $\mu$Hz and $\langle D_{\nu}\rangle_{\mathrm{B}}=0.7$ $\mu$Hz
for $\alpha$~Cen~A and B) have been taken into account. These two figures show the good agreement between our model of the $\alpha$ Cen system
and the asteroseismic observations.
The theoretical frequencies for $\alpha$~Cen A and B are given
in Table~\ref{tab:freq}.

\begin{table*}
\caption{Low degree p--mode frequencies (in $\mu$Hz) for the best model (model M2) of $\alpha$~Cen~A and B. The observations are from Bouchy \& Carrier (\cite{bo02}) for
$\alpha$~Cen~A and from Carrier \& Bourban (\cite{ca03}) for $\alpha$~Cen B.}
\begin{center}
\begin{tabular}{cc|ccc|cccc}
\hline
\hline
 & & \multicolumn{3}{|c|}{Observations} & \multicolumn{4}{c}{Model M2}  \\
 & $n$ & \multicolumn{1}{|c}{$l=0$} & \multicolumn{1}{c}{$l=1$} & \multicolumn{1}{c}{$l=2$} & \multicolumn{1}{|c}{$l=0$} & \multicolumn{1}{c}{$l=1$} & \multicolumn{1}{c}{$l=2$} & \multicolumn{1}{c}{$l=3$} \\ \hline
 & 15 & 	& 	 	& 1833.1 & 1717.8 & 1764.7	      & 1814.0	 & 1857.8 \\ 	      
 & 16 & 1841.3 	& 1887.4 	& 1934.9 & 1820.7 & 1868.4	      & 1917.9	 & 1962.0 \\
 & 17 & 	& 1991.7 	& 2041.5 & 1924.2 & 1972.4	      & 2022.9	 & 2067.7 \\
 & 18 &		& 2095.6	& 2146.0 & 2029.1 & 2077.7	      & 2128.1	 & 2173.6 \\
 & 19 & 2152.9	& 2202.8	& 2251.4 & 2134.0 & 2183.4	      & 2233.8	 & 2279.1 \\
$\alpha$~Cen~A & 20 & 2258.4	& 2309.1	& 2358.4 & 2239.3 & 2288.5	      & 2339.4	 & 2385.2 \\
 & 21 & 2364.2	& 2414.3 	& 2464.1 & 2344.6 & 2394.2	      & 2444.8	 & 2491.2 \\ 
 & 22 & 2470.0	& 2519.3	& 2568.5 & 2449.7 & 2500.0	      & 2551.0	 & 2597.4 \\
 & 23 & 2573.1	& 2625.6	& 	 & 2555.5 & 2605.8	      & 2657.3	 & 2704.3 \\
 & 24 &	2679.8  & 2733.2	& 2782.9 & 2661.5 & 2712.4	      & 2763.6	 & 2810.9 \\
 & 25 &	2786.2	& 2837.6	& 2887.7 & 2767.5 & 2818.6	      & 2870.3	 & 2917.5 \\ 										      
\hline
 & 17 & 	& 3052.0	& 	 & 2977.2 & 3053.2	      & 3125.6	 & 3194.2 \\ 	      
 & 18 &  	& 	 	& 	 & 3136.7 & 3213.2	      & 3286.2	 & 3355.1 \\
 & 19 & 	& 	 	& 	 & 3296.9 & 3373.4	      & 3446.8	 & 3516.3 \\
 & 20 &		& 3532.9	& 3607.2 & 3457.2 & 3534.0	      & 3607.5	 & 3677.4 \\
 & 21 & 3617.2	& 		& 	 & 3617.6 & 3694.5	      & 3768.5	 & 3838.8 \\
$\alpha$~Cen B  & 22 & 	& 3855.0	& 3930.6 & 3778.2 & 3855.4	      & 3929.8	 & 4000.7 \\ 
 & 23 & 3937.6	& 4016.3 	& 	 & 3939.2 & 4016.8	      & 4091.5	 & 4163.0 \\ 
 & 24 & 	& 4177.4	& 4253.8 & 4100.6 & 4178.6	      & 4253.7	 & 4325.4 \\
 & 25 & 	& 		& 	 & 4262.6 & 4340.6	      & 4415.9	 & 4488.2 \\
 & 26 &	4423.1  & 		& 	 & 4424.6 & 4502.8	      & 4578.3	 & 4650.7 \\
 & 27 &	4585.0	& 		& 	 & 4586.7 & 4665.0	      & 4740.8	 & 4813.5 \\     
\hline
\label{tab:freq}
\end{tabular}
\end{center}
\end{table*}

\section{Comparison with previous studies}
 
We initially consider the calibrations made without seismic constraints and in particular
the two recent detailed studies by Guenther \& Demarque (\cite{gu00}) and Morel et al. (\cite{mo00}).
Contrary to our calibration, Guenther \& Demarque (\cite{gu00}) did not use the masses of Pourbaix et al. (\cite{po02})
which were of course not available at the time. However, they considered three different parallaxes which led to
masses for the $\alpha$~Cen system very close to the ones of Pourbaix et al. (\cite{po02}). Indeed, they used masses
contained between $1.0844$ and $1.1238$\,$M_\odot$ for $\alpha$~Cen~A and $0.9017$ and $0.9344$\,$M_\odot$ for
$\alpha$~Cen~B (see Sect.~\ref{intro} for more details). Given that these masses are similar to the ones we used for our
calibration and that the input physics of both evolution codes is similar (mixing--length theory
for convection and helium and heavy--element diffusion), one expects good agreement between their results and ours.
This is indeed the case.
First of all, the initial helium aboundance of our models ($Y_{\mathrm{i}}=0.275 \pm 0.010$) is in good agreement with the
value of about $0.28$ obtained by Guenther \& Demarque. Secondly, Guenther \& Demarque found an age of 
$6.8 \pm0.8$\,Gyr if $\alpha$~Cen~A does not have a convective core. As explained in Sect.~\ref{modM2}, our model of
$\alpha$~Cen~A lies very close to the boundary between models with and without a convective core, but does not exhibit a
convective core. Thus, the age of $6.52 \pm 0.30$\,Gyr of our M2 model is perfectly compatible with the results of 
Guenther \& Demarque. Finally, Guenther \& Demarque found that the mixing--length parameter of
$\alpha$~Cen~A is about 10\,\% smaller that the one of the B component. However, they pointed out that this difference 
may not be significant, because the observables available at the time of their studies were not sufficient to
strongly constrain the models.     
Thanks to the asteroseismic results of Bouchy \& Carrier (\cite{bo02}) and Carrier \& Bourban (\cite{ca03}), 
we obtained a model of the $\alpha$~Cen system which is now firmly constrained. 
As a result, we confirm that a single mixing--length parameter cannot be found for both components of the system, as
already suggested by many authors (Noels et al. \cite{no91}, Fernandes \& Neuforge \cite{fe95} and Guenther \& Demarque \cite{gu00}).
Indeed, our results show that this parameter is about $8$\,\% larger for $\alpha$~Cen
B than for $\alpha$~Cen~A, in perfect agreement with the value of about 10\,\% determined by Guenther \& Demarque. 
Note that, at first sight, we would expect this parameter, if physically meaningful, to be the same for 
all stars regardless of mass, composition and age. However, as shown by models with rotational mixing or mixing by
magnetic instability, many more effects act, to the first order, as a change of the mixing--length parameter. Thus, differences in
$\alpha \equiv l/H_{\mathrm{p}}$ must rather be considered as a measure of our ignorance in the internal stellar hydrodynamics,
rather than only a difference {\it stricto sensu} of the mixing--length parameter.
Note also that the mixing--length parameter of the A component is only slightly larger than the solar
calibrated mixing--length parameter ($\alpha_{\odot}=1.7$). 

The calibration of Morel et al. (\cite{mo00}) was based on the masses obtained by Pourbaix et al. (\cite{po99}):
$M_{\mathrm{A}}=1.16 \pm 0.031$\,$M_\odot$ and $M_{\mathrm{B}}=0.97 \pm 0.032$\,$M_\odot$. These masses are larger than
the ones considered by Guenther \& Demarque (\cite{gu00}) and the new masses of Pourbaix et al. (\cite{po02}) that we used for our
models. Accordingly, the results obtained by Morel et al. (\cite{mo00})
are quite different. For example, Morel et al. found an age of only
$2.71$\,Gyr (for their model with the mixing--length theory and without overshooting). They also found values of
the convection parameter that are almost equal for both stars. All these differences are mainly due to the different masses used. Note also that
Morel et al. used the spectroscopic gravities to constrain the models instead of the luminosities derived from the
photometry. This results in different error boxes in the HR diagram that have to be matched by the models
and thus also influences the results of the calibration.       

The first study of the $\alpha$~Cen system which took into account the asteroseismic measurements of the A component
(Bouchy \& Carrier \cite{bo02}) was made by Th\'evenin et al. (\cite{th02}). At first sight, one can think that our
results are incompatible with this study, since Th\'evenin et al. pointed out that they were unable to produce a model of 
$\alpha$~Cen~A and B compatible with the non--asteroseismic constraints and the data of Bouchy \& Carrier (\cite{bo02})
using the masses of Pourbaix et al. (\cite{po02}). To understand why we are able to determine a model of the $\alpha$~Cen
system which is in good agreement with all observables now available for these stars, one has to compare the
non--asteroseismic constraints used by the two groups. As explained in Sect.~\ref{tchim}, we adopted the same effective
temperature as Th\'evenin et al. for $\alpha$~Cen~B, while we increased the uncertainty on the effective temperature of  
$\alpha$~Cen~A ($5810 \pm 50$\,K) in order to encompass the intervals of temperature used by Th\'evenin et al.
and Thoul et al (\cite{th03}). In the same way, we adopted 
$[\mathrm{Fe/H}]_{\mathrm{A}}=0.22 \pm 0.05$\,dex for $\alpha$~Cen~A and $[\mathrm{Fe/H}]_{\mathrm{B}}=0.24 \pm 0.05$\,dex for 
$\alpha$~Cen~B; these values lie between the ones used by Th\'evenin et al. and those adopted by Thoul et al, with
larger error boxes which appear to us more realistic in view of the different values found in the literature. Finally,
Th\'evenin et al. determined the luminosities of both stars from the Geneva photometry using mean values with errors which
do not take into account the quality of the night. We also used the Geneva photometric 
system, but calculated the mean magnitudes using the individual values listed in Table~\ref{tab:phot} and 
taking into account the quality of the night. As a result, the luminosities used by Th\'evenin et al. are very similar
to our luminosities but with smaller uncertainties. It is important to underline these small differences, because they can explain the
different results found by the two studies. Indeed, our model of the $\alpha$~Cen system gives a luminosity of
$\alpha$~Cen~B which is in good agreement with our adopted observational constraints, but not with the small error box in
the HR diagram adopted by Th\'evenin et al. Consequently, it is not surprising that Th\'evenin et al. did not find the
solution we obtained.
Moreover, the input physics of the evolution code used by Th\'evenin et al. differs from the input physics of our code. The main
difference concerns the modelisation of convection: Th\'evenin et al. used the Canuto and Mazzitelli formulation while
we used the mixing--length theory. These two different formulations can lead to significant different results, as shown by
Morel et al. (\cite{mo00}). Another important difference concerns the hypothesis of a unique convection parameter for both stars. Indeed, Th\'evenin et al. assumed the
convection parameters to be the same for both components. As aforesaid, we found that a single mixing--length 
parameter is not applicable for both components of the system, in good agreement with Th\'evenin et al. 

We also notice that the
model proposed by Th\'evenin et al. is not in good agreement with the new asteroseismic measurements of $\alpha$~Cen~B (Carrier \& Bourban \cite{ca03}) due to theoretical large and small spacings that are higher than observed. Besides,
Kervella et al. (\cite{ke03}) slightly changed the masse and the mixing--length parameter of the B component in order to
obtain a model of $\alpha$~Cen~B with a slightly larger radius than the one found by Th\'evenin et al. which is 
compatible with the interferometric radius. This larger radius for $\alpha$~Cen~B certainly
improves the agreement between their model and the observed mean large spacing. It is perhaps possible
to obtain a model based on the masses derived by Th\'evenin et al. which is also compatible with the observed mean small spacing 
by redoing the whole calibration in the same way as Th\'evenin et al., but with the asteroseismic measurements 
of $\alpha$~Cen~B as additional constraints. However, we consider the observed masses as true observables
which have to be reproduced by a consistent model of the $\alpha$~Cen system.  

Finally, we compare our results to the second calibration of $\alpha$~Cen A and B which took into account the
asteroseismic observations of $\alpha$~Cen~A (Thoul et al. \cite{th03}). 
At first sight, one can think that the study by Thoul et al. is in good agreement with our work, since they determined a
model of the $\alpha$~Cen system which reproduced the observed p--mode frequencies as well as the masses of Pourbaix et al.
(\cite{po02}). Moreover, they found an age of $6.41$\,Gyr which is in good agreement with our value of $6.52 \pm 0.30$\,Gyr.
However, we have to notice that the non--asteroseismic observational constraints they used are quite different from ours.
Indeed, Thoul et al. used the spectroscopic surface gravities, while we used the luminosities derived from the photometry,
bolometric correction and parallax. We chose to use the luminosities instead of the surface gravities to constrain the
models, because the luminosities are determined with higher accuracy than the surface gravities. Consequently, our error
boxes in the HR diagram are smaller than the ones adopted by Thoul et al. As a result, we see that our model of
$\alpha$~Cen~B lies within the error box adopted by Thoul et al., whereas the model of $\alpha$~Cen~B determined
by Thoul et al. is not included in our error boxes; this simply means that our model is also in perfect agreement with the
surface gravities, whereas the solution obtained by Thoul et al. is not compatible with the observed luminosity of
$\alpha$~Cen B. Another main difference between both studies concerns the input physics of the
evolution code. Indeed, the evolution code used for our calibration includes a detailed treatment of the 
atomic diffusion of helium and heavy elements, contrary to Thoul et al. who neglect it. This fact can also
explain certain discrepancies between the results found by Thoul et al. and our results, and in particular the fact
that they obtained an identical mixing--length parameter for both stars.

\section{Conclusion}
              
The aim of this work was to determine the best model for the $\alpha$~Cen system using the Geneva evolution code including
atomic diffusion. This model had to reproduce all
observational constraints available for $\alpha$~Cen~A and B, namely the masses, the luminosities, the effective temperatures, 
the metallicities, the radii and the low degree p--mode frequencies of both stars.
First, we used all non--asteroseismic constraints with one--sigma error boxes and found the solution $t=6.50 \pm 0.20$\,Gyr, $\alpha_{\mathrm{A}}=1.83 \pm 0.10$, $\alpha_{\mathrm{B}}=1.99 \pm 0.10$, 
$Y_{\mathrm{i}}=0.275 \pm 0.010$ and $(Z/X)_{\mathrm{i}}=0.0435 \pm 0.0020$. However, this model, which is in perfect
agreement with all non--asteroseismic measurements, was found to have large spacings $\Delta \nu_0$ slightly larger than
observed ($105.9 \pm 0.1$\,$\mu$Hz instead of $105.5 \pm 0.1$\,$\mu$Hz for $\alpha$~Cen~A 
and $161.7 \pm 0.1$\,$\mu$Hz instead of $161.1 \pm 0.1$\,$\mu$Hz for $\alpha$~Cen B). These small discrepancies indicate that
the radii observed by interferometry are slightly smaller than the ones deduced from asteroseismic measurements.        
Accordingly, we increased the error boxes on the observed radii from one to two sigma and found the
following solution: $t=6.52 \pm 0.30$\,Gyr, $\alpha_{\mathrm{A}}=1.83 \pm 0.10$, $\alpha_{\mathrm{B}}=1.97 \pm 0.10$, 
$Y_{\mathrm{i}}=0.275 \pm 0.010$ and $(Z/X)_{\mathrm{i}}=0.0434 \pm 0.0020$. 
Note that the radii of this solution are still in good agreement
with the interferometric results of Kervella et al. (\cite{ke03}) since they are only $0.4$\,\% and $0.8$\,\% larger than the
observed radii for $\alpha$~Cen~A and B respectively. 	      
	      
To sum up, we point out that we obtained a model of the $\alpha$~Cen system which correctly reproduce all 
the numerous observational constraints now available for both stars. Thanks to the recent asteroseismic measurements
of both components (Bouchy \& Carrier \cite{bo02} and Carrier \& Bourban \cite{ca03}), this model is firmly
constrained.   
We therefore conclude that asteroseismic measurements are needed to determine accurate stellar parameters of a given star
as well as to test the physics used in the stellar evolution codes. Indeed, non--asteroseismic constraints are not able to
discriminate stellar models with similar positions in the HR diagram but different internal structures.
It is however worthwhile to recall that asteroseismic measurements by themselves are not sufficient to obtain a reliable
model of a given star nor to test the physics of stellar models. Indeed, the analysis of $\alpha$~Cen~A and B shows that
the combination of asteroseismic and non--asteroseismic constraints is the only way to obtain interesting and accurate
results. Consequently, the determination of precise non--asteroseismic parameters like the luminosity, the effective temperature
or the metallicity of a given star is needed.     
For this purpose, the analysis of a visual binary system constitutes an ideal target for future asteroseismic
measurements, since it is the guarantee to obtain reliable non--asteroseismic parameters and additional constraints as the same
initial chemical composition and age for both stars.  

\begin{acknowledgements}
We would like to thank J. Christensen--Dalsgaard for providing us with the Aarhus adiabatic pulsation code.
We also thank D. Pourbaix for helpful advices.
This work was partly supported by the Swiss National Science Foundation. 
\end{acknowledgements}


\begin{thebibliography}{}
\bibitem[1999]{an99} Angulo, C., et al. 1999, Nucl. Phys. A, 656, 3 
\bibitem[2002]{bo02} Bouchy, F., \& Carrier, F. 2002, A\&A, 390, 205
\bibitem[1979]{burnet} Burnet, M., \& Rufener, F. 1979, A\&A, 74, 54
\bibitem[1991]{ca91} Canuto, V.M., \& Mazzitelli, I. 1991, ApJ, 370, 295
\bibitem[1992]{ca92} Canuto, V.M., \& Mazzitelli, I. 1992, ApJ, 389, 724
\bibitem[2003]{ca03} Carrier, F., \& Bourban, G. 2003, A\&A, 406, L23 
\bibitem[1997]{cd97} Christensen--Dalsgaard, J. 1997,\\
$\mathtt{http://www.obs.aau.dk/\! \sim \! jcd/adipack.n/}$
\bibitem[1988]{da88} D\"appen, W., Mihalas, D., Hummer, D.G., \& Mihalas, B.W. 1988, ApJ, 332, 261
\bibitem[1986]{de86} Demarque, P., Guenther, D.B., \& van Altena, W.F. 1986, ApJ, 300, 773
\bibitem[1992]{ed92} Edmonds, P., Cram, L., Demarque, P., et al. 1992, ApJ, 394, 313
\bibitem[2001]{en01} Endl, M., K\"urster, M., Els, S., Hatzes, A.P., \& Cochran, W.D. 2001, A\&A, 374, 675
\bibitem[1995]{fe95} Fernandes, J., \& Neuforge, C. 1995, A\&A, 295, 678
\bibitem[1978]{fl78} Flannery, B.P., \& Ayres, T.R. 1978, ApJ, 221, 175
\bibitem[1996]{flower} Flower, P. 1996, ApJ, 469, 355
\bibitem[1984]{fo84} Fossat, E., Grec, G., Gelly, B., \& Decanini, Y. 1984, {\it Comptes Rendus Acad. Sci.
Paris}, S\'erie 2, 229, 17
\bibitem[1980]{golay} Golay, M. 1980, Vistas in Astronomy 24, 141
\bibitem[1998]{gr98} Grevesse, N., \& Sauval, A.J. 1998, Space Sci. Rev., 85, 161
\bibitem[2000]{gu00} Guenther, D.B., \& Demarque, P. 2000, ApJ, 531, 503
\bibitem[1982]{he82} Heintz, W.D. 1982, The Observatory, 102, 42
\bibitem[1988]{hu88} Hummer, D.G., \& Mihalas, D. 1988, ApJ, 331, 794
\bibitem[1978]{ka78} Kamper, K.W., \& Wesselink, A.J. 1978, AJ, 83, 1653 
\bibitem[2003]{ke03} Kervella, P., Th\'evenin, F., S\'egransan, D., et al. 2003, A\&A, 404, 1087
\bibitem[1999]{ki99} Kim, Y.-C. 1999, JKAS, 32, 119
\bibitem[1998]{le98} Lejeune, T., Cuisinier, F., \& Buser, R. 1998, A\&AS, 130, 65
\bibitem[1993]{ly93} Lydon, T.J., Fox, P.A., \& Sofia, S. 1993, ApJ, 413, 390
\bibitem[2000]{mm00} Meynet, G., \& Maeder, A. 2000, A\&A, 361, 101
\bibitem[1988]{mi88} Mihalas, D., D\"appen, W., Hummer, D.G. 1988, ApJ, 331, 815
\bibitem[2000]{mo00} Morel, P., Provost, J., Lebreton, Y., \& Berthomieu, G. 2000, A\&A, 363, 675
\bibitem[1993]{ne93} Neuforge, C. 1992, A\&A, 268, 650
\bibitem[1997]{ne97} Neuforge--Verheecke, C., \& Magain, P. 1997, A\&A, 328, 261
\bibitem[1991]{no91} Noels, A., Grevesse, N., Magain, P., et al. 1991, A\&A, 247, 91
\bibitem[1986]{pa86} Paquette, C., Pelletier, C., Fontaine, G., \& Michaud, G. 1986, ApJ, 61, 177
\bibitem[1999]{po99} Pourbaix, D., Neuforge--Verheecke, C. \& Noels, A. 1999, A\&A, 344, 172
\bibitem[2002]{po02} Pourbaix, D., Nidever, D., McCarthy, C., et al. 2002, A\&A, 386, 280
\bibitem[1996]{ri96} Richard, O., Vauclair, S., Charbonnel, C., \& Dziembowski, W.A. 1996, 312, 1000
\bibitem[1964]{rufener1} Rufener, F. 1964, Publ. Obs. Gen\`eve, A, 66, 413
\bibitem[1985]{rufener2} Rufener, F. 1985, in \textit{Calibration of
Fundamental Stellar Quantities}, IAU Symp. 111, 253 (Eds. D.S. Hayes et al)
Reidel Publ. Co., Dordrecht
\bibitem[1999]{so99} S\"oderhjelm, S. 1999, A\&A, 341, 121
\bibitem[1980]{ta80} Tassoul, M. 1980, AJS, 43, 469
\bibitem[2002]{th02} Th\'evenin, F., Provost, J., Morel, P., et al. 2002, A\&A, 392, L9
\bibitem[2003]{th03} Thoul, A., Scuflaire, R., Noels, A., et al. 2003, A\&A, 402, 293
\bibitem[1998]{tu98} Turcotte, S., Richer, J., Michaud, G., et al. 1998, ApJ, 504, 539
\end{thebibliography}
\end{document}